\documentclass[structabstract]{raa}

\usepackage{soul}
\usepackage{color}
\usepackage{booktabs}
\usepackage{graphicx,times}
\usepackage{natbib,amssymb}
\usepackage{epstopdf,epsfig}
\usepackage{float}
\usepackage{amsmath}
\usepackage{ulem}
\usepackage{amsmath} 
\usepackage{hyperref}
\usepackage{mathtools}

\providecommand{\bm}[1]{\mbox{\boldmath$#1$\unboldmath}}

\begin{document}
 \title{Ultra-compact X-ray Binaries: A Review}

   \volnopage{ {\bf 2026} Vol.\ {\bf X} No. {\bf XX}, 000--000}
   \setcounter{page}{1}
   \author{Bo Wang\inst{1}, 
 	       Dongdong Liu\inst{1},
   	       Yunlang Guo\inst{2},
           Hailiang Chen\inst{1},
           Wen-Cong Chen\inst{3},
           \and
           Zhanwen Han\inst{1}
          }
   \institute{International Center of Supernovae (ICESUN), Yunnan Key Laboratory of Supernova Research, Yunnan Observatories, Chinese Academy of Sciences, Kunming 650216, China;
   {\it wangbo@ynao.ac.cn; liudongdong@ynao.ac.cn; yunlang@nju.edu.cn}\\
          \and
          School of Astronomy and Space Science, Nanjing University, Nanjing 210023, China\\
          \and
          School of Science, Qingdao University of Technology, Qingdao 266525, China\\
              }

\date{Received ; accepted}

\abstract {
Ultra-compact X-ray binaries (UCXBs) are a subclass of low-mass X-ray binaries (LMXBs) characterized by ultra-short orbital periods, typically less than $60-80$\,min. They consist of a compact mass-accretor and a hydrogen-poor mass-donor, in which the mass-accretor could be a neutron star (NS) or even a black hole (BH).
UCXBs play an important role in multiple areas of astrophysics. In particular, they are considered strong, continuous gravitational wave (GW) sources in the low-frequency band, making them key targets for future space-based GW observatories such as LISA, TianQin and Taiji. As the most compact binaries, the formation and evolution of UCXBs remain highly uncertain.
In this article, we summarize theoretical advances and observational status in UCXBs.
Especially, we review four classic formation channels: 
the white dwarf donor channel, the He star donor channel, the evolved main-sequence donor channel, and the accretion-induced collapse channel. 
We also discuss recent progress in these channels, 
covering evolutionary scenarios, the initial parameter space of the progenitors, and associated objects. 
A comparison between observations and theoretical predictions is provided, 
along with a discussion on the observed BH-UCXB candidates and their formation path.
The origin of UCXBs can be constrained by the chemical composition of mass-donors and their locations in diagrams of mass-transfer rate (or X-ray luminosity) versus orbital period. We also examine the implications of UCXBs for several astrophysical fields, including GW astronomy, multi-messenger astronomy, binary evolution, and NS physics under extreme conditions. Further progress will depend on multi-wavelength observations, the discovery of more samples, and more detailed theoretical simulations.
 }

\titlerunning{UCXBs}

\authorrunning{B. Wang et al.}

\maketitle

\section{Introduction}

Low-mass X-ray binaries (LMXBs) are a type of mass-accretion powered X-ray sources, 
in which the mass-accretor could be a neutron star (NS) or a stellar-mass black hole (BH) 
that is accreting matter from a low-mass ($<1~M_\odot$) companion through
Roche-lobe overflow (RLOF; see \citealt{Bhattacharya1991PhR...203....1B,
Tauris1999A&A...350..928T,Podsiadlowski2002ApJ...565.1107P}). Binary millisecond pulsars (MSP) were generally thought to evolve from NS-LMXBs, in which a NS is spun up to millisecond periods by accreting matter and angular momentum from a low-mass donor  (see, e.g., \citealt{Alpar1982Natur.300..728A,Bhattacharya1991PhR...203....1B}). A few hundred LMXBs have been observed in the Galaxy, containing both transients and persistent sources (e.g., \citealt{LiuQZ2007A&A...469..807L,Casares2017hsn..book.1499C}). 

Ultra-compact X-ray binaries (UCXBs) are a subclass of LMXBs with ultra-short orbital
periods that are usually less than $60-80$\,min, consisting of a compact mass-accretor 
and a hydrogen-poor mass-donor (see \citealt{Rappaport1982ApJ...254..616R,Savonije1986AA...155...51S,Nelson1986ApJ...304..231N,
Nelemans2010NewAR..54...87N}).
Another kind of interacting ultra-compact binary is AM CVn systems, which 
have similar mass-donors as UCXBs but the mass-accretors are white dwarfs 
(WDs; see \citealt{Warner1995Ap&SS.225..249W,Podsiadlowski2003MNRAS.340.1214P,
Nelemans2010NewAR..54...87N,Solheim2010PASP..122.1133S,liu21,ChenHaiLiang2022ApJ...935....9C,Green2025AA...700A.107G,YuW.2026A&A...706A..14Y}). 

UCXBs play a key role in modern astrophysics, as follows: 
(1)  UCXBs are important low-frequency gravitational wave (GW) sources due to their short orbital periods, 
which can be detected by space-based GW detectors, such as LISA (e.g., 
\citealt{Nelemans2009CQGra..26i4030N, Amaro-Seoane2017arXiv170200786A,Tauris2018PhRvL.121m1105T,Amaro-Seoane2023LRR....26....2A}),
TianQin  (e.g., \citealt{Luo2016CQGra..33c5010L, Bao2019PhRvD.100h4024B,Huang2020PhRvD.102f3021H, 
LiEn-Kun2025RPPh...88e6901L}),
and Taiji (e.g., \citealt{Ruan2020NatAs...4..108R, Ruan2021Resea202114164R}).  
Meanwhile, UCXBs are also dual-band GW sources that can be potentially used to constrain the properties of NSs 
(see, e.g., \citealt{Bildsten1998ApJ...501L..89B,Suvorov2021MNRAS.503.5495S,ChenWen-Cong2021PhRvD.103j3004C}).
(2) 
In addition to GW radiation,
UCXBs also emit X-ray and optical radiation simultaneously, and especially exhibit radio radiation in their late stages. This allows for simultaneous multi-wavelength observations, which is of great significance for the study of multi-messenger astronomy.
(3) UCXBs are important astrophysical labs owing to the combination of compact mass-accretors, 
ultra-short orbits and  mass-donors with different chemical compositions  (see
 \citealt{Nelemans2010NewAR..54...87N}).
(4) It has been proposed that UCXBs are possible progenitor candidates of MSPs (see, e.g., \citealt{Alpar1982Natur.300..728A,YangZ.L.2023ApJ...956L..39Y,GuoYunlang2025ApJ...992..144G}).
(5) UCXBs provide important constraints on the binary evolution theory under extreme conditions, 
e.g., the common-envelope (CE) evolution, the angular-momentum loss mechanisms, 
 the accretion physics of compact objects, etc (see \citealt{Tauris2023pbse.book.....T}).

The observations of UCXBs can be traced back to the 1970s, and the 
number of UCXB candidates increased significantly as more X-ray satellites were launched after the 1990s, such as the ROSAT observatory, the Chandra X-ray Observatory, 
the XMM-Newton mission, the Rossi X-ray Timing Explorer (RXTE), etc
(see \citealt{ArmasPadilla2023AA...677A.186A}).
Up to now, there are about 50 UCXB candidates, in which 49 sources 
have been discovered in the Galaxy and one is in M31 
(see \citealt{ArmasPadilla2023AA...677A.186A,zhangJiachang2024MNRAS.530.2096Z, 
Moutard2024ApJ...968...51M,MaQian-Qi2026}).
In the Galaxy, 
half of UCXB candidates are located in the bulge, 
and $\sim90\%$ sources are located within 10\,kpc of the Sun (see \citealt{ArmasPadilla2023AA...677A.186A}).
\citet{ArmasPadilla2023AA...677A.186A} 
suggested that there are at least two distinct classes for UCXB candidates, as follows:
(1) Persistent systems with orbital periods $<30$\,min account for the UCXB population in globular clusters.
(2) Transients (70\%) with  orbital periods between 40 and 60\,min are dominated by sources formed in the Galactic field.

In Table 1, we list the observational properties of 22 confirmed UCXBs  
that have been identified with high confidence based on their precise 
measurements for orbital periods, including 9 transients and 13 persistent sources  (see, e.g., \citealt{intZand2007AA...465..953I,Heinke2013ApJ...768..184H,
Cartwright2013ApJ...768..183C,
ArmasPadilla2023AA...677A.186A,MaQian-Qi2026}). 
In the Galaxy, there are 21 confirmed UCXBs, in which 9 in the field, 
8 in globular clusters and 4 in the bulge.
So far, more than half of the observed UCXBs have been discovered in globular clusters or in the bulge.
Meanwhile, about half of the confirmed UCXBs are the accreting MSPs (see \citealt{ArmasPadilla2023AA...677A.186A}).

In the observations,
most of currently confirmed UCXBs have been suggested to contain NS accretors though 
they are not identified exclusively   
(see, e.g., \citealt{Sazonov2020NewAR..8801536S,CotiZelati2021AA...650A..69C}). 
It is worth noting that three BH-UCXB candidates have been suggested,
i.e., 47 Tuc X9 (see \citealt{Bahramian2017MNRAS.467.2199B,Tudor2018MNRAS.476.1889T}), 
RZ 2109 (see \citealt{Steele2014ApJ...785..147S,Dage2024MNRAS.529.1347D})
and RX\,J0042.3$+$4115 (i.e., M31 UCXB-1; see \citealt{MaQian-Qi2026}). 
For more discussions on BH-UCXBs, see Sect. 6.

Due to the ultra-short orbits, the mass-donors in UCXBs could be constrained to be hydrogen-poor,   
fully or partially degenerate stars  
(e.g., WDs or He stars; see \citealt{Rappaport1982ApJ...254..616R,Nelson1986ApJ...304..231N,Deloye2003ApJ...598.1217D}).
According to the spectra in some UCXBs, it has been suggested  that 
the accreted material onto NSs is composed of He or C and O,  
which can be used to identify the mass-donors in UCXBs
 (see \citealt{Nelemans2004MNRAS.348L...7N,Nelemans2006MNRAS.370..255N,vanHaaften2012A&A...543A.121V}). 
In addition, \citet{Galloway2008ApJS..179..360G} suggested that thermonuclear (type I) X-ray bursts 
from accreting NSs have distinct properties independent of the accreted material,  
which may help to constrain  the compositions of the mass-donors in UCXBs
(see also \citealt{Cumming2003ApJ...595.1077C}).

In dense star clusters (e.g., globular clusters), UCXBs  are believed to result from dynamical interactions (tidal captures or stellar encounters) with a high probability
(see, e.g., \citealt{Verbunt1987ApJ...312L..23V,Bailyn1987ApJ...316L..25B,Davies1992ApJ...401..246D,
Davies1998MNRAS.301...15D,Ivanova2005ApJ...621L.109I,Verbunt2005AIPC..797...30V,Voss2007MNRAS.380.1685V}). 
In the Galactic field, however, their formation is dominated by isolated binary evolution. 
It has been proposed that the  mass-donors in UCXBs could be a WD
(the WD donor channel), a He star  (the He star donor channel) or an evolved  main-sequence (MS) star (the evolved MS donor channel; for a review see \citealt{Nelemans2010NewAR..54...87N}).
Meanwhile, an UCXB can also be formed through 
the accretion-induced collapse (AIC) of WDs in binaries, known as the AIC channel (see \citealt{LiuD.2023MNRAS.521.6053L}).

In this review, we mainly focus on the formation and evolution of UCXBs.
In Sects 2-5, we review the WD donor channel, the He star donor channel, the evolved MS donor channel and the AIC channel, respectively. In Sect. 6,
we discuss the observed BH-UCXB candidates and the formation channels to BH-UCXBs.  
In Sect. 7, 
we make a comparison between theoretical results and observations.
The impacts of UCXBs on some research fields are discussed in Sect. 8.
A summary and perspective are provided in Sect. 9.

\begin{table*}[h!]
 \begin{tabular}{c@{}c@{}c@{}c@{}c@{  }c@{}c@{}c@{}c@{  }c }
 	\toprule
	\hline
  Object & $P_{\rm orb}$ \,\,& $P_{\rm spin}$  &$F_{\rm X}$[2-10\,keV] &$\dot M_{\rm tr}$ & $M_{\rm c}$ & \,\,Composition & $d$ & Location & Ref.\\
    & (min)&(ms) & (erg\,cm$^{-2}$\,s$^{-1}$)&($M_\odot$/yr) &($M_\odot$)  & &(kpc)& \\
  \hline
  
  \multicolumn{9}{c}{Transient sources} \\
  
  \hline
  XTE J1807-294 &$40.1$ &$5.25$&$1.19\times10^{-10}$ &$<1.9^{+2.6}_{-1.6}\times10^{-11}$&$0.022$ & -&$8^{+4}_{-3.3}$& Bulge&1,2,3\\
 XTE J1751-305 & $42.4$&$2.30$&$1.27\times10^{-9}$& $5.1^{+2.6}_{-2.9}\times10^{-12}$&$0.02$&-& 
 $8^{+0.5}_{-1.3}$& Bulge&1,2,3\\
  XTE J0929-314 & $43.6$&$5.40$&$2.7\times10^{-10}$& $<9.7^{+25}_{-7.7}\times10^{-12}$&$0.01$&He,C,O&$8^{+7}_{-3}$& Field&1,2,3,4\\
  MAXI J0911-655 & $44.3$&$2.94$&$1.37\times10^{-10}$& $(0.5-3.8)\times10^{-10}$&$0.028$&-& $10.1^{+0.112}_{-0.111}$& NGC 2808&1,5\\
  IGR J16597-3704 & $46.0$&$9.51$&$1.97\times10^{-10}$& $5.5\times10^{-10}$&$0.01$&- & $7.2^{+0.299}_{-0.288}$ & NGC 6256&1,6\\
 Swift J1756.9-2508& $54.7$&$5.49$&$1.54\times10^{-9}$& $1.9^{+2.5}_{-1.7}\times10^{-11}$&$0.015$ &He&$8^{+4}_{-4}$& Bulge&1,2,3,4\\
 NGC 6440 X-2& 57.3&$4.86$&$1.58\times10^{-10}$&$1.3\pm0.7\times10^{-12}$& $0.008$ &-& $8.2^{+0.248}_{-0.241}$&NGC 6440&1,2,3\\
 MAXI J1957+032&$60.9$&$3.18$&$4.91\times10^{-10}$&$0.8\times10^{-10}$&$0.085$&- & $5\pm2$& Field&1,7\\
 IGR J17494-3030& $75.0$&$2.66$&$9.35\times10^{-11}$&$2.6\times10^{-12}$ &$0.02$ &-&8?& Field&1,8\\
  \hline %

    \multicolumn{9}{c}{Persistent sources} \\
    
  \hline
  RX J0042.3+4115 & 7.7 &?&$1.36\times10^{-12}$&-&0.088&-&785&M31&9\\
  4U 1728-34  &10.8?&-&$2.08\times10^{-9}$& $2.6\pm1.6\times10^{-9}$&-&-& $3.3\pm0.5$&Bulge&1,2\\
  4U 1820-303 & $11.4$ &-&$5.02\times10^{-9}$& $1.2\pm0.8\times10^{-8}$&$0.07$&He& $8.0\pm0.108$&NGC 6624    &1,2,3,4\\
  4U 0513-40  & $17.0$&-&$1.33\times10^{-10}$& $1.2\pm0.6\times10^{-9}$&$0.05$&He&$12.0^{+0.134}_{-0.133}$& NGC 1851&1,2,3,4\\
  2S 0918-549 & $17.4$ &-&$1.12\times10^{-10}$&$2.6\pm1.5\times10^{-10}$&-&He,C,O?& $3.0\pm0.2$&Field&1,2,4\\
  4U 1543-624 & $18.2$ &-&$7.27\times10^{-10}$& $1.3^{+1.8}_{-1.2}\times10^{-9}$&$0.03$&C,O&$9.2\pm2.3$& Field &1,2,3,4\\
 4U 1850-087 & $20.6$&-&$1.04\times10^{-10}$&$2.8\pm1.4\times10^{-10}$&$0.04$&Ne-excess& $7.4^{+0.240}_{-0.233}$&NGC 6712&1,2,3,4\\
  M15 X-2    &22.6 &- &$6.71\times10^{-11}$& $3.8\pm1.9\times10^{-10}$&0.03&He?,C?& $10.7\pm0.095$ &NGC 7078&1,2,3,4 \\
  47 Tuc X9 & $28.2$&?&$4.73\times10^{-12}$& $(0.3-1)\times10^{-11}$ &$0.015$&C,O& $4.5\pm0.031$&NGC 104&1,10\\
 IGR J17062-6143 &$38.0$&6.11&$5.53\times10^{-11}$&$1.8^{+1.8}_{-0.5}\times10^{-10}$ &0.01&O&$7.3\pm0.5$& Field&1,3,11\\
 4U 1626-67 & $41.5$ &7680&$2.69\times10^{-10}$&$8^{+14}_{-6}\times10^{-10}$&$0.04$&C,O,Ne&$8$ &Field&1,2,3,4\\
 4U 1916-053 & $50.0$& - &$5.57\times10^{-10}$&$6.3\pm3.7\times10^{-10}$&$0.012$&He,N & $5.8\pm0.3$ & Field&1,2,3,4\\
 4U 0614+091 & $51.3$ &   - &$1.77\times10^{-9}$&$3.9\pm2.3\times10^{-10}$ &$0.014$&C,O & $2.0\pm0.02$ &Field&1,2,3,4\\
  \hline
 \end{tabular}
\caption{Catalogue of 22 confirmed UCXBs, listing key parameters including the orbital period ($P_{\rm orb}$), the spin period ($P_{\rm spin}$), the unabsorbed peak X-ray flux ($F_{\rm X}$[2-10\,keV]), the estimated average mass-transfer rate ($\dot M_{\rm tr}$), the companion mass ($M_{\rm c}$), the composition, the distance ($d$) and the location. The sample is divided into transients (top) and persistent sources (bottom).  \\
\textbf{References:} 1. \citet{ArmasPadilla2023AA...677A.186A}; 2. \citet{Heinke2013ApJ...768..184H}; 3. \citet{2023A&A...675A.199A};
 4. \citet{vanHaaften2012A&A...543A.121V}; 5. \citet{2017A&A...598A..34S};
 6. \citet{2018A&A...610L...2S}; 7. \citet{2022MNRAS.516L..76S};
 8. \citet{2021ApJ...908L..15N}; 9. \citet{MaQian-Qi2026}; 10. \citet{Tudor2018MNRAS.476.1889T};
 11. \citet{2019MNRAS.488.4596H}.
}
\label{all_sources}
 \end{table*}

\section{The WD donor channel}\label{sect:The WD donor channel}

\subsection{Formation and evolution of NS+He WD systems}

\begin{figure}
\begin{center}
\epsfig{file=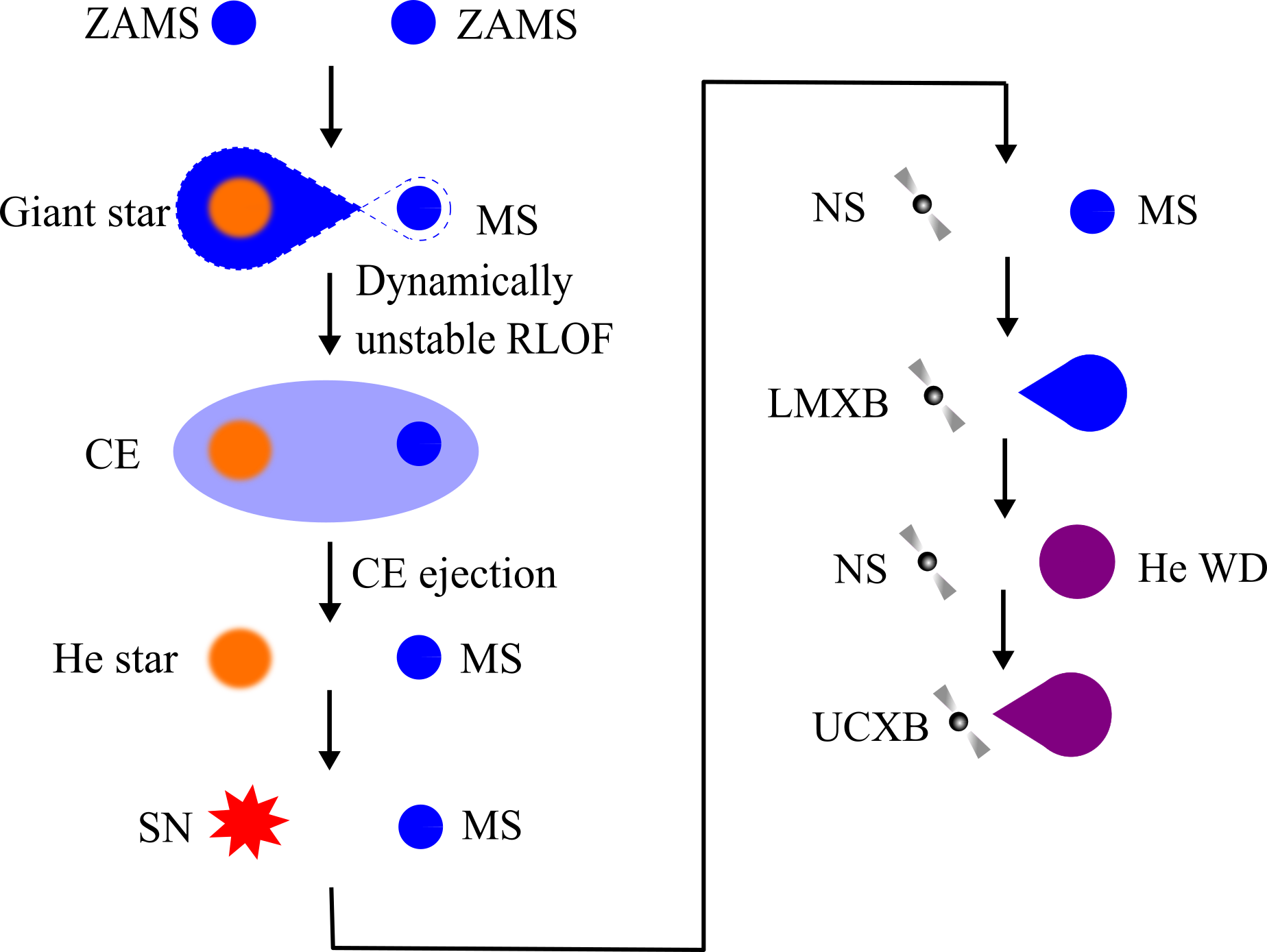,angle=0,width=11.cm}
\caption{Formation of NS+He WD systems that can evolve into UCXBs.}
\label{fig:NS+HeWDchannel}
    \end{center}
\end{figure}

In this channel, 
a NS accretes He-rich material from a He WD through Roche-lobe overflow (RLOF), 
in which the mass-transfer is driven by GW radiation
(see, e.g., \citealt{Pringle1975MNRAS.172..493P,Iben1995ApJS..100..233I,
Belczynski2004ApJ...603..690B,
vanHaaften2012A&A...537A.104V,YuShenghua2021MNRAS.503.2776Y,ChenHai-Liang2022ApJ...930..134C}). 
There is one evolutionary scenario to form NS+He WD systems and then produce UCXBs (see Fig. 1).
In the primordial binary, the primary star evolves faster onto the giant branch stage
and begins transferring material onto its companion. 
Due to the large mass-ratio, the mass-transfer process becomes dynamically unstable, 
leading the binary into a CE stage. During this stage, 
the companion star orbits the core of the primary, and their orbital energy is converted into the internal energy of the envelope. If sufficient orbital energy is available, the CE can be ejected, resulting in the formation of a He star+MS binary. 
The He star then evolves and collapses into a NS via a supernova (SN) explosion. If the system survives, a NS+MS binary is formed. Subsequently, magnetic braking (MB) causes the donor star to fill its Roche lobe near the end of the MS or during the Hertzsprung gap (HG) phase, initiating mass-transfer onto the NS. Once this mass-transfer concludes, a MSP+He WD binary is produced. At this stage, the He WD is extremely low-mass and the orbital period is relatively short. 
GW radiation then causes the orbit to shrink until the He WD fills its Roche lobe. The system subsequently evolves into an UCXB, eventually forming an isolated MSP or a pulsar+planet-like system.
For more studies on the formation and evolution of NS$+$WD systems, see, e.g., 
\citet{LvGuoliang2017ApJ...847...62L},
\citet{Toonen2018A&A...619A..53T},
\citet{GuWei-Min2020MNRAS.497.1543G},
\citet{LiZhenwei2024ResPh..5907568L},
\citet{YangXing-Peng2025ApJ...995...99Y},
\citet{LiKayeJiale2025ApJ...984...41L}
and
\citet{KangYacheng2026A&A...706A.106K}, etc.

\subsection{Parameter space for UCXBs}

\begin{figure}
\begin{center}
\epsfig{file=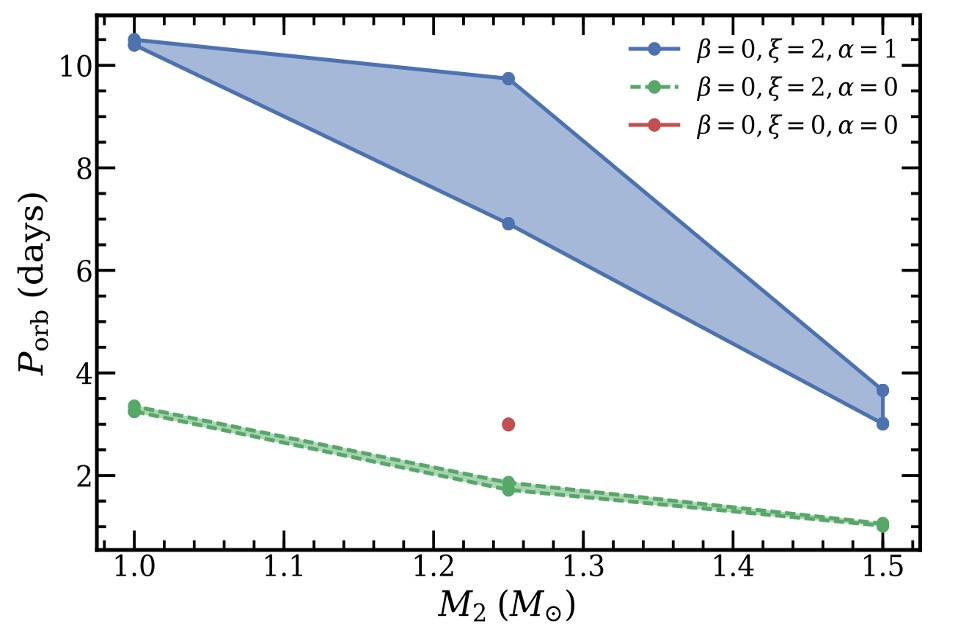,angle=0,width=11.cm}
\caption{Parameter space for producing UCXBs through the WD donor channel, 
in which the initial NS mass is assumed to be $1.3\,M_{\odot}$. 
It has been suggested that
models with  $\beta=0$, $\xi= 2$ and  $\alpha= 1$ can better 
reproduce the properties of the observed X-ray binaries
(see \citealt{Van2019MNRAS.483.5595V}).
Source: From \citet{ChenHaiLiang2021MNRAS.503.3540C}.}
  \end{center}
  \label{fig:WDchannel}
\end{figure}

By modelling the formation of UCXBs  
with the standard MB prescription from \citet{Rappaport1983ApJ...275..713R}, 
previous studies argued that the parameter space from the WD donor channel is relatively small (see \citealt{van2005A&A...431..647V,van2005A&A...440..973V}). 
Meanwhile, it has been suggested that it is impossible to 
produce UCXBs if a weaker MB is adopted (see \citealt{van2005A&A...431..647V,van2005A&A...440..973V}). 
In addition, \citet{Istrate2014A&A...571A..45I} suggested that 
the parameter space of MSPs with extremely low-mass WD companions
is also very small if the standard MB prescription is adopted. 

By using a newly suggested MB prescription from \citet{Van2019MNRAS.483.5595V},
\citet{ChenHaiLiang2021MNRAS.503.3540C} recently found that the initial orbital period range of LMXBs forming UCXBs becomes significantly wider (see Fig. 2; see also \citealt{DengZhuLing2021ApJ...909..174D}). 
These studies suggested that a stronger MB is needed for modelling the formation of UCXBs. 
\citet{YangXing-Peng2025ApJ...995...99Y} arrived at a similar conclusion by modeling 
the formation of MSP+He WDs with a different MB prescription.

The stability of mass-transfer from a WD onto a NS is still being debated. 
In earlier studies, 
a critical WD mass is used to determine the stability of mass-transfer, in which  
the critical mass ranges from $0.35$ to $1.25\;M_{\odot}$. 
In most of these studies, a semi-analytic method is adopted to model the evolution of NS+WD binaries
(see, e.g., \citealt{1988ApJ...332..193V,2009PhRvD..80b4006P,Yungelson2002A&A...388..546Y,vanHaaften2012A&A...537A.104V,YuShenghua2021MNRAS.503.2776Y}). 
According to the results of hydrodynamic simulations, \citet{Bobrick2017MNRAS.467.3556B} 
argued that only binaries including He WDs with masses 
$<0.2\,M_{\odot}$ can undergo stable mass-transfer and then produce UCXBs.
In addition, by considering the detailed structure of He WDs, 
\citet{ChenHai-Liang2022ApJ...930..134C} modelled the evolution of NS+He WD binaries with detailed binary evolution computations, suggesting that the critical WD mass is $0.45\,M_{\odot}$.
The main reason for the difference between the two studies is that \citet{Bobrick2017MNRAS.467.3556B} included the effect of disk wind, which causes extra angular momentum loss and thereby leads to a smaller critical mass.  
Previous studies indicate that it is hard to produce UCXBs with CO WD donors (see, e.g., \citealt{Bobrick2017MNRAS.467.3556B, ChenHai-Liang2022ApJ...930..134C}). Moreover, \citet{Sengar2017MNRAS.470L...6S} suggested that the WD donor channel can well explain the observed properties of UCXBs with high He abundances.

These UCXBs, along with detached NS+He WD binaries that have short orbital periods, can generate strong GW emission in the mHz band, detectable by LISA, TianQin and Taiji. However, the number of detectable sources remains highly uncertain.
\citet{HeJian-Guo2024MNRAS.529.1886H} modelled the Galactic population of NS+CO/ONe WD systems, predicting that approximately tens to hundreds of such systems could be detected by LISA. In a complementary estimate, \citet{Tauris2018PhRvL.121m1105T} suggested that LISA could detect at least 100 UCXBs with NS accretors.

Additionally, 4U 1820-303 is one of the most compact binaries, which has a negative orbital-period derivative 
in globular cluster NGC 6624 (see, e.g., \citealt{Sansom1989PASJ...41..591S,ChouY2001ApJ...563..934C,Peuten2014ApJ...795..116P}).
According to the WD donor channel,
\citet{JiangLong2017ApJ...837...64J} used an evolutionary circumbinary disk assumption to account for the negative orbital-period derivative of this source.
However, \citet{ChouY2001ApJ...563..934C} argued that 4U 1820-303 may originate from
a hierarchical triple system with a $\sim$\,1.1\,d period companion based on its stable long-term modulation.

\section{The He star donor channel}\label{sect:The He star donor channel}

\begin{figure}
\begin{center}
\epsfig{file=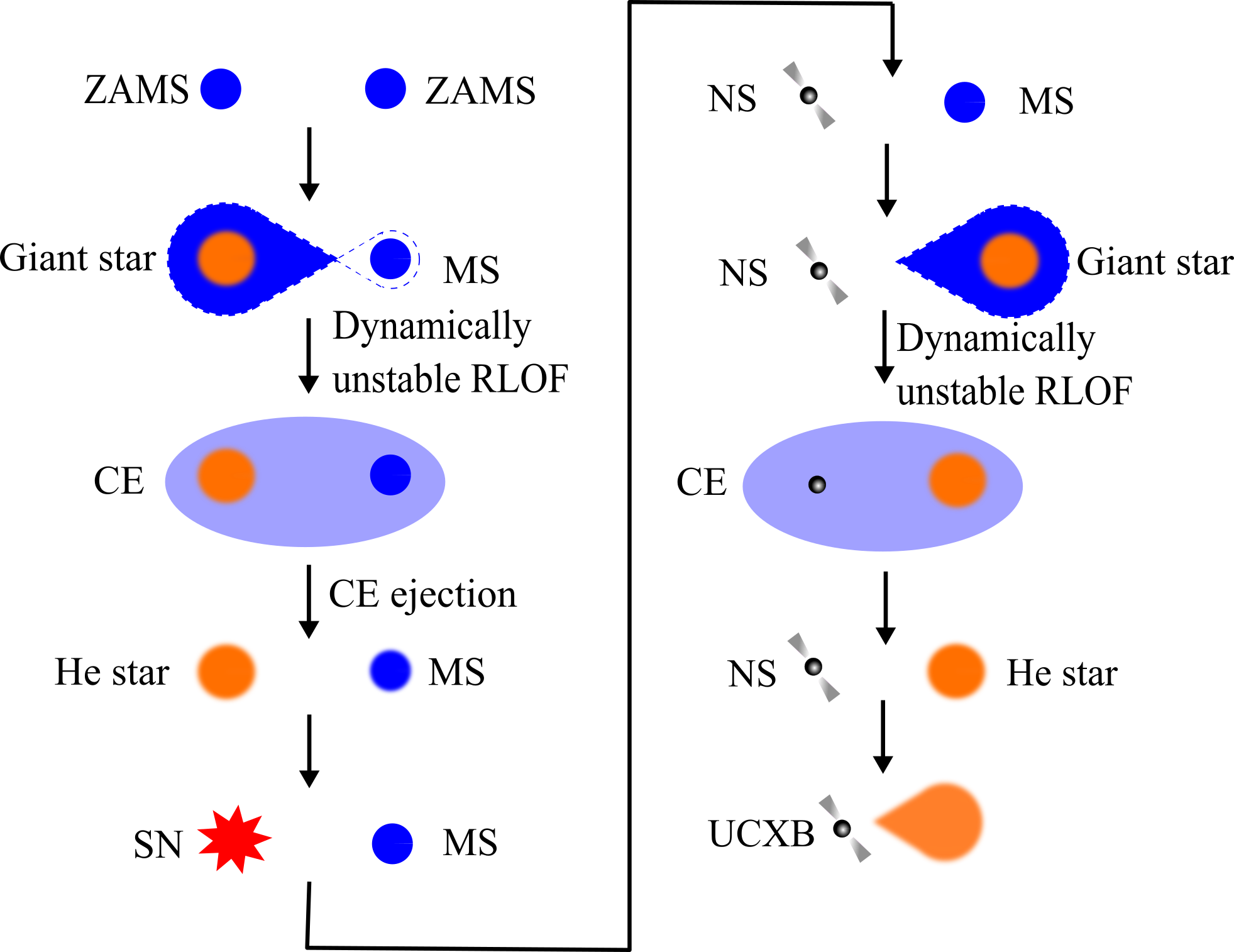,angle=0,width=11.cm}
\caption{Formation of NS+He star systems that can evolve into UCXBs.}
  \end{center}
  \label{fig:Hestarchannel}
\end{figure}

\subsection{Formation and evolution of NS+He star systems}

In this channel,
a NS accretes matter from a He star through RLOF, 
in which the He star fills its Roche-lobe due to the rapid orbital shrinkage induced 
by GW radiation
(see, e.g., \citealt{Savonije1986AA...155...51S,Heinke2013ApJ...768..184H,Wang2021MNRAS.506.4654W}).
There is one evolutionary scenario to form NS+He star systems and then produce UCXBs (see Fig. 3). 
The primordial primary first fills its Roche lobe during its subgiant phase, initiating a dynamically 
unstable mass-transfer because of its large mass-ratio.
In this case, a CE would be formed.
If the CE can be ejected, the primordial primary will evolve into a He star.
The He star continues to evolve and will explode as an SN, leading to the formation of a NS.
Subsequently, the primordial secondary fills its Roche-lobe when it evolves to the giant stage, leading to the formation of the second CE.
After the ejection of the CE, a NS$+$He star system would be produced.
The He star will fill its Roche-lobe due to the contraction of the orbit resulting from the GW radiation.
In this case, the mass-transfer is stable and the system evolves into an UCXB (see \citealt{Wang2021MNRAS.506.4654W}).
For more studies on the formation and evolution of NS$+$He star systems, see, e.g., 
\citet{GuoYunlang2023MNRAS.526..932G},
\citet{LiLuhan2024MNRAS.534.3400L},
\citet{Guo2024MNRAS.530.4461G},
\citet{YangZ.L.2025Sci...388..859Y},
\citet{DengZhu-Ling2025A&A...704A...2D} and
\citet{Benvenuto2026A&A...705L...6B}, etc.

\subsection{Parameter space for UCXBs}

\begin{figure}
\begin{center}
\epsfig{file=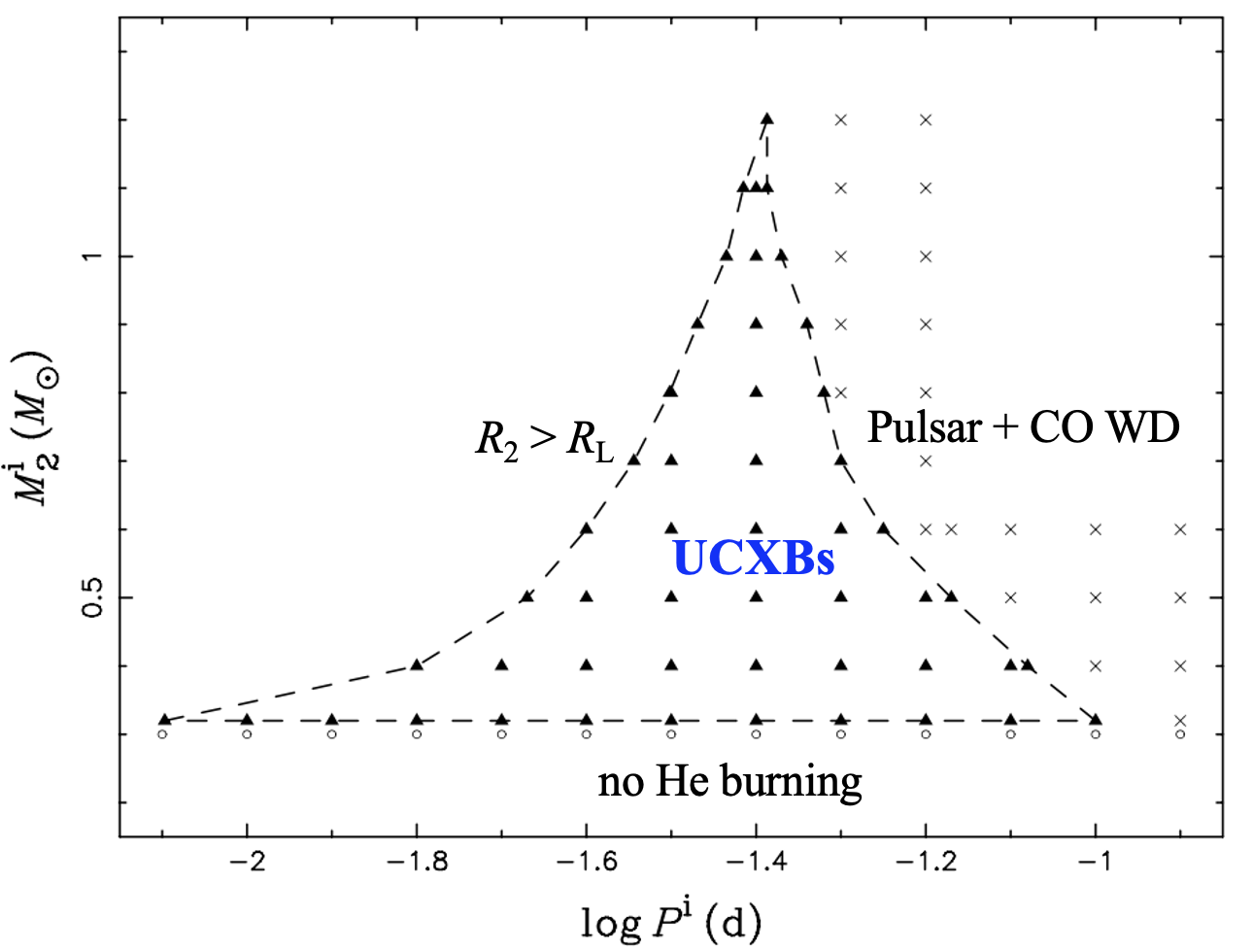,angle=0,width=11.cm}
\caption{Parameter space for producing UCXBs through the He star donor channel,
in which the initial NS mass is assumed to be $1.4\,M_{\odot}$. 
The filled triangles are for those resulting in the formation of UCXBs. 
Open circles are those that the masses of He star donors are too low to ignite He in their center. 
Crosses indicate systems that will experience high mass-transfer rates 
when the He star donors evolve into the subgiant stage, 
leading to the formation of intermediate-mass binary pulsars. 
Source: From \citet{Wang2021MNRAS.506.4654W}.}
  \end{center}
  \label{fig:channel}
\end{figure}

By computing the evolution of a NS+He star system with a tight orbit,
\citet{Savonije1986AA...155...51S} first confirmed that NSs can be accompanied by He stars
with low masses.
It has been suggested that 50\%$-$80\% of UCXBs might be produced through the He star donor channel (see \citealt{ZhuChun-Hua2012RAA....12.1526Z}).
\citet{Wang2021MNRAS.506.4654W} recently followed the long-term evolution of NS+He star systems, thereby obtaining the initial parameter space for the formation of UCXBs (see Fig. 4). 
The parameter space in Fig. 4 is set by the following criteria:
(1) The lower boundary of the contour is determined by the  condition that
the masses of  He stars  should be larger than $0.32\, M_{\odot}$,
below which He burning  would be extinguished in their center (see also \citealt{Han2002MNRAS.336..449H}).
(2) The left boundary is constrained by the
requirement that the He star fills its Roche-lobe
during the zero-age MS phase.
(3) The systems beyond the right boundary 
undergo high mass-transfer rates when the He stars evolve into the subgiant stage, 
forming intermediate-mass binary pulsars.

\citet{Wang2021MNRAS.506.4654W} 
found that all NS+He star systems forming UCXBs can be detectable by the LISA within a distance of 15\,kpc.
These NS+He star systems are even be visible as LISA sources within a distance of 100\,kpc. 
The average timescale that the NS+He star binary appears to be as a LISA source is about 33\,Myr within a distance of 15\,kpc.
By using a detailed binary population synthesis method (see \citealt{HanZhan-Wen2020RAA....20..161H}), \citet{Wang2021MNRAS.506.4654W} 
estimated that the Galactic rates of UCXB-LISA sources  are $\sim3.1-11.9\, \rm Myr^{-1}$, and that
the number of UCXB-LISA sources can be $\sim1-26$ through the calibration of UCXB observations in the Galaxy.

\subsection{Black widows with extremely low masses}

Black widows (BWs) are a subclass of eclipsing MSPs in compact binaries with low-mass companions
(see, e.g., \citealt{2013IAUS..291..127R, ChenHaiLiang2013ApJ...775...27C, Benvenuto2014ApJ...786L...7B}).
Their companions are generally thought to be gradually stripped by irradiation-driven winds powered by $\gamma$-ray emission and relativistic particles from the MSP, a mechanism commonly referred to as the evaporation process
(e.g., \citealt{Kluzniak1988Natur.334..225K, Ruderman1989ApJ...336..507R}).
BWs provide valuable laboratories for investigating the evolutionary pathways of close NS binaries,
including the magnetic-field decay, the angular-momentum loss, the accretion efficiency, and the constraints on the NS equation of state, etc
(e.g., \citealt{Cumming2001ApJ...557..958C, Tauris2012MNRAS.425.1601T, Van2019MNRAS.483.5595V}).

\begin{figure}
\begin{center}
\epsfig{file=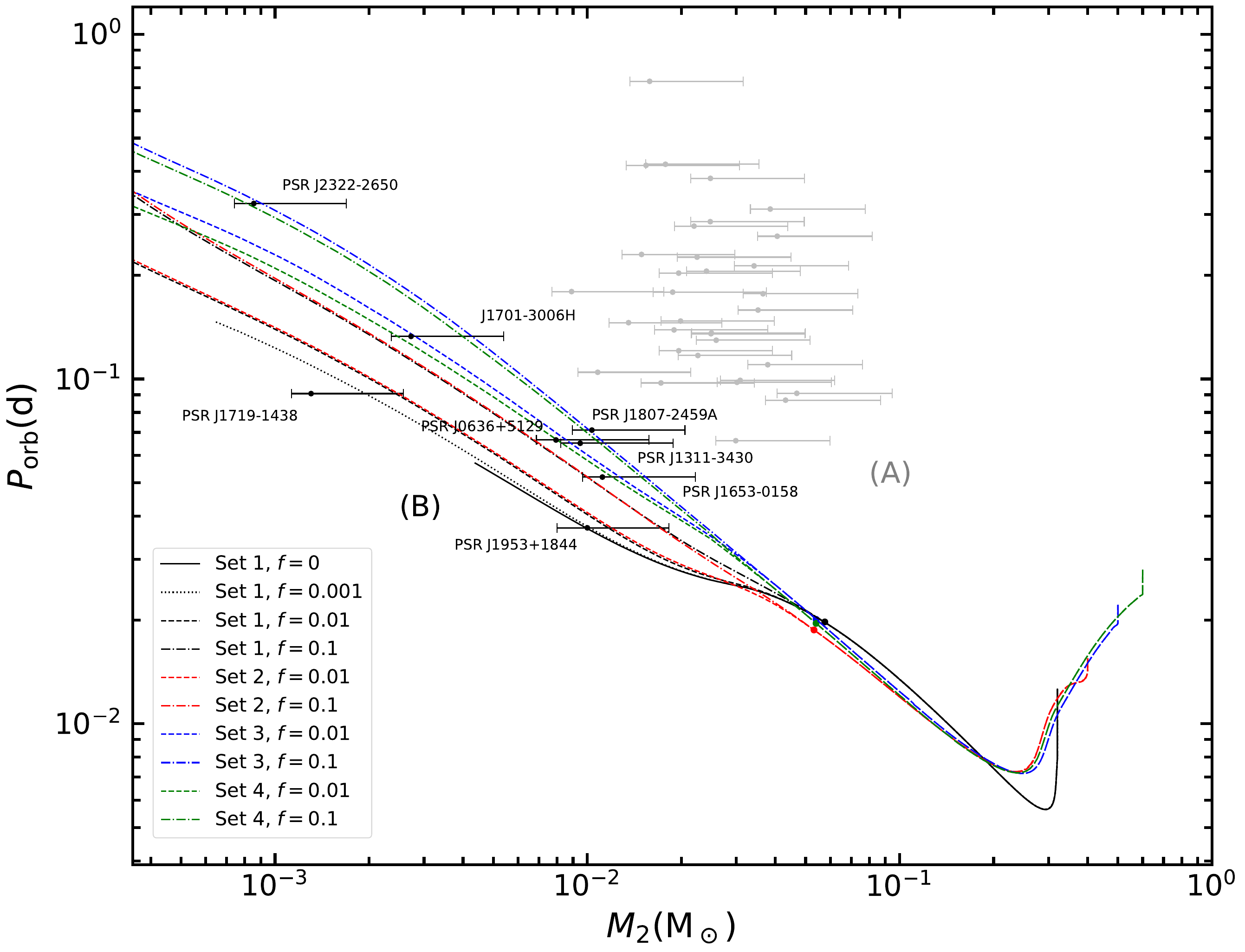,angle=0,width=11.cm}
\caption{The evolutionary tracks of NS+He star binaries 
with different initial He star masses ($M_2$) and evaporation efficiencies ($f$), along which the systems may evolve through the UCXB phase.
Different colors represent the NS+He star systems with different initial parameters,
hereafter referred to as sets 1–4 (see also \citealt{GuoYun-Lang2022MNRAS.515.2725G}). Set 1 starts with $M_2^{\rm i}=0.32\,M_\odot$ and $\log (P_{\rm orb}^{\rm i}/\rm d)=-1.90$, while sets 2–4 correspond to progressively higher initial He star masses ($M_2^{\rm i}=0.40$, 0.50, and $0.60\,M_\odot$) and wider initial orbits ($\log (P_{\rm orb}^{\rm i}/\rm d)=-1.80$, $-1.65$, and $-1.55$).
Colored dots mark the onset of the evaporation process for each set.
Black symbols in region (B) indicate BWs with $M_2 < 0.01\,M_\odot$, while gray symbols in region (A) represent BWs with $M_2$ in the range $0.01$–$0.05\,M_\odot$.
Observational data for BWs are taken from the ATNF Pulsar Catalogue: \href{http://www.atnf.csiro.au/research/pulsar/psrcat}{http://www.atnf.csiro.au/research/pulsar/psrcat}
(version 1.70, 2023 May; \citealt{Manchester2005AJ....129.1993M}).
The error bars indicate the minimum and maximum companion masses,
corresponding to orbital inclination angles of $90^{\circ}$ and $26^{\circ}$ (the 90\% probability limit), respectively. Compared with Fig. 4 in \citet{GuoYun-Lang2022MNRAS.515.2725G}, we added a new BW in this figure, i.e., PSR J1953+1844.}
  \end{center}
  \label{fig:bw}
\end{figure}

With the growing number of BW systems identified,
it has become increasingly evident that BWs can be divided into two subpopulations based on their companion masses ($M_2$).
Fig.\,5 shows the distribution of BWs in the $P_{\rm orb}-M_2$ plane.
Systems located in region (A) correspond to BWs with $M_2$ in the range of $0.01-0.05\,M_\odot$, while those in region (B) have extremely low-mass companions with $M_2\lesssim0.01\,M_\odot$
(see \citealt{GuoYun-Lang2022MNRAS.515.2725G}).
The origin of BWs in region (B) cannot be explained by conventional evolutionary scenarios that fail to reduce the companion mass to such small values within the Hubble time
(see, e.g., \citealt{ChenHaiLiang2013ApJ...775...27C, Ginzburg2020MNRAS.495.3656G}).

Based on the He star donor channel for the formation of UCXBs
(see \citealt{Wang2021MNRAS.506.4654W}),
\citet{GuoYun-Lang2022MNRAS.515.2725G} considered the evaporation process into the subsequent evolution.
Their models successfully reproduce the observed properties of BWs with extremely low-mass companions ($M_2\lesssim0.01\,M_\odot$),
including PSRs J1719-1438, J2322-2650 and J1653-0158, etc.
More recently,
the Five-hundred-meter Aperture Spherical radio Telescope (FAST) discovered a binary pulsar  PSR J1953+1844 (i.e., M71E) in a 53\,min orbit, which is
currently the shortest orbital-period binary pulsar known
(see, e.g., \citealt{HanJL2021RAA....21..107H, PanZ2023Natur.620..961P}).
This system can also be naturally explained through the UCXBs with the He star donor channel
(e.g., \citealt{YangZ.L.2023ApJ...956L..39Y, Guo2024MNRAS.527.7394G}).
In addition,
\citet{Vleeschower2024MNRAS.530.1436V} recently reported the discovery of 
a MSP J1701-3006H (i.e., M62H) in the globular cluster M62 using MeerKAT,
hosting an extremely low-mass companion with a median mass of $\sim0.0027\,M_\odot$ and an orbital period of 2.7\,h.
Such a low companion mass for M62H indicates an evolutionary origin 
through the He star donor channel (see Fig. 5).

BWs have long been proposed as a possible evolutionary channel leading to isolated MSPs
(e.g., \citealt{Kluzniak1988Natur.334..225K, vandenHeuvel1988Natur.334..227V}).
However,
the standard channel fails to account for the observed isolated MSP population within the Hubble time
(e.g., \citealt{vanHaaften2012A&A...541A..22V, Ginzburg2020MNRAS.495.3656G}).
The simulations by \citet{GuoYun-Lang2022MNRAS.515.2725G} indicate that
the companion mass can be reduced to $<10^{-3}\,M_\odot$, or even $<10^{-5}\,M_\odot$.
Therefore,
the BWs formed through the  He star donor channel  are promising progenitors for isolated MSPs.

\section{The evolved MS donor channel}
\label{sect:The evolved MS donor channel}

\subsection{Formation and evolution of NS+MS systems}

In this channel, a NS accretes matter from an evolved  MS donor with initial orbital periods below the bifurcation period, eventually evolving into UCXBs
(see \citealt{Podsiadlowski2002ApJ...565.1107P,ChenWen-Cong2020ApJ...900L...8C,ChenMinghua2025ApJ...981..175C}).
In the classic scenario,
LMXBs with initial orbital periods above the bifurcation period will produce diverging systems, otherwise forming converging systems 
(see, e.g., \citealt{Pylyser1988A&A...191...57P,Pylyser1989A&A...208...52P}).
The evolutionary scenario for this channel is similar to that of the WD donor channel (see Fig.~\ref{fig:NS+HeWDchannel}), 
but the formed NS+MS systems with initial orbital periods should be below the bifurcation period.

Previous studies indicate that 
the bifurcation period is sensitive to the mechanisms of
the angular-momentum loss, especially for the prescriptions of 
the MB laws 
(see, e.g., \citealt{Ergma1996AA...315L..17E,Podsiadlowski2002ApJ...565.1107P,
van2005A&A...431..647V,van2005A&A...440..973V,mabo2009ApJ...691.1611M,Shao2015ApJ...809...99S,ChenWen-Cong2020ApJ...900L...8C,ChenHaiLiang2021MNRAS.503.3540C,DengZhuLing2021ApJ...909..174D}). 
Some NS+MS binaries can evolve into detached MSP+WD systems if their initial orbital periods are very close to the bifurcation periods (see \citealt{ChenWen-Cong2020ApJ...900L...8C}). Subsequent evolution overlaps with the WD donor channel. In the standard MB case, it requires fine-tuning initial orbital periods to form MSP+WD systems with short orbital periods (WDs can fill its Roche lobe due to GW radiation in the Hubble time) for those NS+MS binaries (see \citealt{Istrate2014A&A...571A..45I}). Otherwise, NS+MS systems always experience a mass-transfer rather than a detached stage even if their initial orbital periods are shorter than the bifurcation periods (see \citealt{ChenWen-Cong2020ApJ...900L...8C}). Some efficient MB mechanisms can alleviate the fine-tuning problem and widen the range of initial orbital periods that can form detached MSP+WD systems (see, e.g., \citealt{ChenHaiLiang2021MNRAS.503.3540C,YangXing-Peng2025ApJ...995...99Y}). 

For intermediate-mass X-ray binaries (IMXBs, donor-star masses $\ge 2~M_\odot$), the mass-transfer occurs on a short thermal timescale because the material is transferred from the more massive donor star to the less massive NS. In principle, IMXBs spend most of their X-ray active lifetime as LMXBs. Therefore, 
some IMXBs can firstly evolve into LMXBs and then evolve toward UCXBs (see, e.g.,\citealt{Podsiadlowski2002ApJ...565.1107P,LiXiangDong2002ApJ...564..930L,XuXiaoJie2007AA...476.1283X,Lin2011ApJ...732...70L,ShaoYong2012ApJ...756...85S, Li2020RAA....20..162L}). 
Because intermediate-mass stars without convective envelopes are not generally expected to experience the MB process, it seems that IMXBs are difficult to evolve into UCXBs. However, previous studies indicate that some pre-IMXBs with initial orbital periods very close to the bifurcation periods can still evolve into UCXBs due to the MB process once donor stars develop convective envelope (i.e., donor-star masses $\le1.5~M_\odot$; see, e.g., \citealt{Istrate2014A&A...571A..45I,ChenWen-Cong2020ApJ...900L...8C}). About 5\% of intermediate-mass stars (so-called Ap/Bp stars) possess anomalously strong magnetic fields, it can drive IMXBs to evolve toward UCXBs
by anomalous MB mechanism induced by the coupling between the irradiation-driven wind and the magnetic field 
(see \citealt{ChenWen-Cong2016ApJ...830..131C}).
Furthermore, the tidal torque between an IMXB and a surrounding circumbinary disk can also efficiently extract angular momentum from the system and induce the formation of an UCXB (see \citealt{MaBo2009ApJ...698.1907M}).

\subsection{Parameter space for UCXBs}
Adopting the standard MB precription given by \cite{Rappaport1983ApJ...275..713R},
\citet{ChenWen-Cong2020ApJ...900L...8C} used the stellar evolution code MESA to simulate
the evolution of a large number of pre-LMXBs/IMXBs containing NS+MS systems. Their results show
that UCXBs could be formed by NS+MS systems with a $0.4-3.5\,M_\odot$ MS companion 
and an initial orbital period shorter than the bifurcation periods  (see Fig. 6). 
As shown in Fig. 6,  most 
pre-LMXBs/IMXBs with initial orbital periods less than 
the bifurcation periods can evolve into UCXB-LISA sources within a distance of 1\,kpc, but the 
parameter space is in a very narrow range within a distance of 10\,kpc. 
According to the calculated parameter space, \citet{ChenWen-Cong2020ApJ...900L...8C} 
estimated 
that the  Galactic rates of UCXB-LISA sources are $\sim2-2.6\,\rm Myr^{-1}$ 
based on binary population synthesis simulations. They obtained  
the number of UCXB-LISA sources in the range of 240-320
by considering the contribution of UCXBs in globular clusters.

\begin{figure}
\begin{center}
\epsfig{file=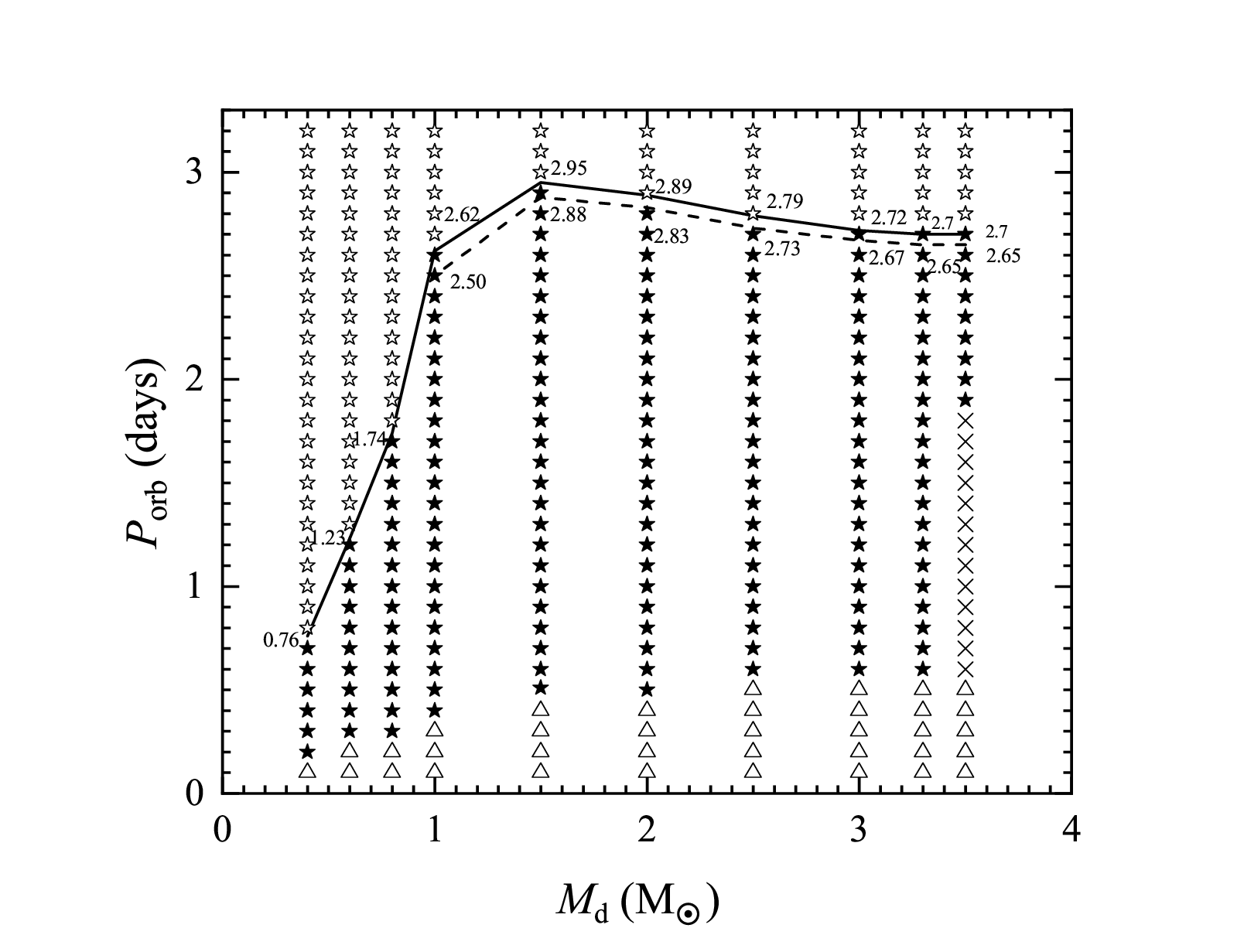,angle=0,width=11cm} 
\caption{Parameter space of the pre-LMXBs/IMXBs for producing UCXBs,
in which the initial NS mass is assumed to be $1.4~M_{\odot}$. 
The solid stars show the pre-LMXBs/IMXBs that can evolve into UCXB-LISA sources 
within a distance of 1\,kpc. 
The solid curve denotes the bifurcation periods of LMXBs/IMXBs with different initial masses of donors,
whereas the dashed curve presents the minimum initial orbital period for the progenitors of UCXB-LISA sources within a distance of 10\,kpc. 
The open stars, the open triangles and the crosses indicate the systems that will not produce UCXBs. 
Source: From \citet{ChenWen-Cong2020ApJ...900L...8C}.} 
\label{fig:mschannel}
\end{center}
\end{figure}

\section{The AIC channel}\label{sect:The AIC channel}
\subsection{Formation and evolution of AIC systems}
In this channel, 
an ONe WD accretes matter from a He WD, eventually collapsing into a NS when the ONe WD increases its mass
close to the Chandrasekhar limit (${M}_{\rm Ch}$; see \citealt{Nomoto+1979wdvd.coll...56N,Taam+1986ApJ...305..235T,Nomoto1991ApJ...367L..19N,Brooks2017ApJ...850..127B}). 
After that,
the NS continues to accrete matter from the He WD, forming an UCXB finally (see \citealt{LiuD.2023MNRAS.521.6053L}). 
This channel is related to the AIC process of ONe WDs in binaries (for a recent review see \citealt{Wang+2020RAA....20..135W}).
AIC events are a type of faint optical transients and 
exhibit rapid changes in brightness, which are important for producing isolated NSs and NS systems  (see, e.g., \citealt{Woosley1992ApJ...391..228W,Fryer1999ApJ...516..892F,Dessart2007ApJ...669..585D}).
Up to date, there is no direct detection for the existence of AIC events though
there exists a lot of indirect evidence in observations 
(see, e.g., \citealt{Canal1990ARA&A..28..183C,Ivanova2008MNRAS.386..553I,Boyles2011ApJ...742...51B,
Tauris2013A&A...558A..39T,Ruiter2019MNRAS.484..698R,Ablimit2019ApJ...881...72A,
Wang2026RAA....26c2001W}).
For more studies on the formation and evolution of
AIC systems, see, e.g., 
\citet{Chen2011MNRAS.410.1441C},
\citet{Brooks2017ApJ...843..151B},
\citet{Wang+2017MNRAS.472.1593W},
\citet{Wang2018MNRAS.481..439W},
\citet{LiuD2018MNRAS.477..384L},
\citet{Wu2019MNRAS.483..263W},
\citet{Liu2020MNRAS.494.3422L},
\citet{Wang2022MNRAS.510.6011W},
\citet{WuChengyuan2023MNRAS.525.6295W},
\citet{Wu2023ApJ...944L..54W},
\citet{ZhangZ2024ApJ...975..186Z}, \citet{Ruiter2025A&ARv..33....1R},
\citet{Batziou2025ApJ...984..197B}
and \citet{MengXiang-Cun2025arXiv250518702M}, etc.

\begin{figure}
\begin{center}
\epsfig{file=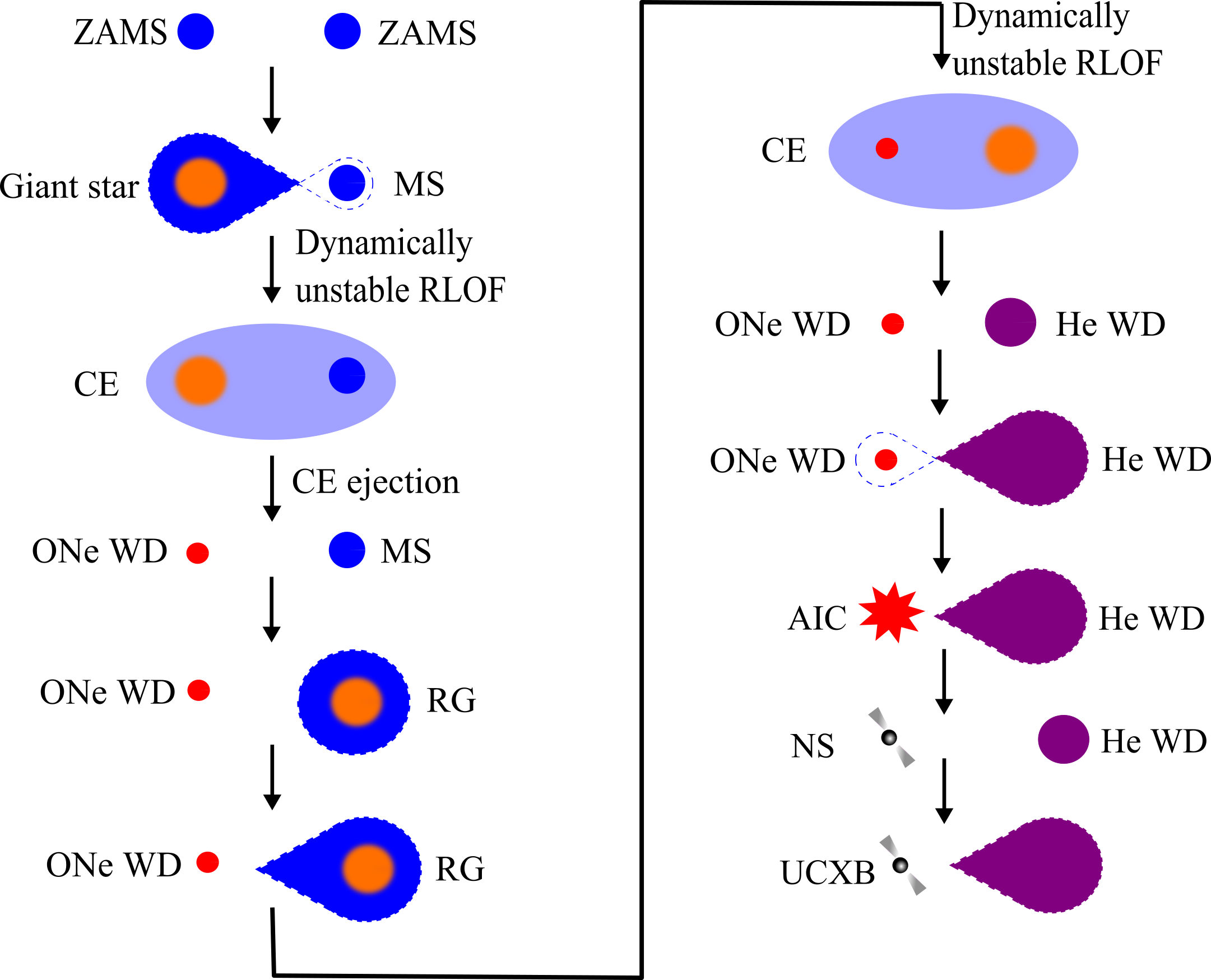,angle=0,width=10.cm}
\caption{Formation of ONe WD+He WD systems that undergo the AIC process and evolve into UCXBs finally.}
  \end{center}
  \label{fig:channel}
\end{figure}

There is one evolutionary scenario to form ONe WD+He WD systems, eventually producing UCXBs through the AIC process (see Fig. 7). 
The primordial primary first fills its Roche lobe during the giant phase, leading to a dynamically unstable mass-transfer process.
In this case, a CE is supposed to be formed.
If the CE can be ejected, an ONe WD$+$MS system would be produced.
Subsequently, the primordial secondary continues to evolve, 
and will fill its Roche-lobe at the red-giant stage.
During this phase, the mass-transfer is also dynamically unstable and a second CE may be formed.
After the second CE ejection, the primordial secondary will evolve into a He WD.
After that, the He WD will fill its Roche-lobe due to the orbital shrinks driven by the GW radiation.
The accreted He-rich matter burns stably on the surface of the ONe WD, leading to the mass-growth of the WD.
When the ONe WD mass increases to ${M}_{\rm Ch}$, an AIC event is supposed to occur, leading to the formation of a NS.
After that, the He WD may fill its Roche-lobe again owing to the GW radiation, leading to the formation of an UCXB (see \citealt{LiuD.2023MNRAS.521.6053L}).

\begin{figure}
\begin{center}
\epsfig{file=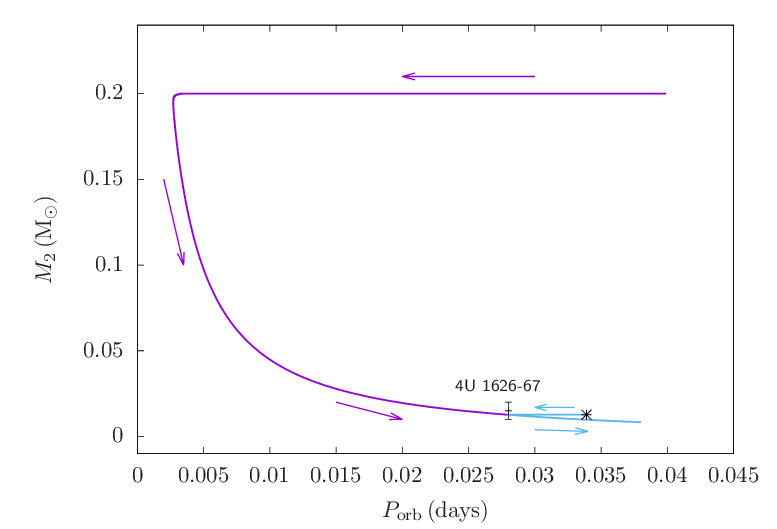,angle=0,width=10.cm}
\caption{An alternative evolutionary track for the formation and evolution of 4U 1626-67. The purple line shows the evolutionary track before the AIC process, 
whereas the blue line presents the evolutionary track after the AIC process.
The snowflake represents the start of the position just after the AIC process. Source: From \citet{LiuD.2023MNRAS.521.6053L}.}
  \end{center}
  \label{fig:channel}
\end{figure}

\subsection{UCXBs from the AIC Channel}
4U 1626–67 is a type of peculiar UCXB, containing a newly formed NS and an ultra-light companion.
However, the evolutionary path for 4U 1626–67 is still under hot debate.
It has been suggested that peculiar newly formed NS$+$ultra-light companion systems 
like 4U 1626–67 could represent evidence for the existence of AIC events (see \citealt{Tauris2013A&A...558A..39T}).
\citet{LiuD.2023MNRAS.521.6053L} carried out a series of ONe WD+He WD systems for producing AIC events, suggesting that the NS binaries would evolve to UCXBs when the He WDs refill their Roche lobes. 
They found that the properties of 4U 1626–67 can be reproduced by the AIC channel. 
Fig. 8 shows a possible evolutionary track for the formation and evolution of 4U 1626-67 
based on the AIC channel.
As shown in this figure, the He WD would refill its Roche-lobe after the AIC process, resulting in the production of an UCXB with a newly formed NS and an ultra-light He WD.
The observed orbital period and companion mass in 4U 1626–67 can be well reproduced by the AIC channel,
although there are still some uncertainties regarding its origin, especially for the high mass-transfer rate (for more discussions see \citealt{LiuD.2023MNRAS.521.6053L}).

In addition, UCXB XTE J1751-305 is a transient accreting MSP with an orbital period of 42\,min (\citealt{Markwardt2002ApJ...575L..21M}).
It has been proposed that  XTE\,J1751$-$305 may also originate from the AIC channel (see \citealt{LiuD.2023MNRAS.521.6053L}).
The asteroseismology models indicate that the accreting matter onto the NS in XTE J1751-305 is composed mostly of He (see \citealt{Lee10.1093/mnras/stu1077, Strohmayer2014ApJ...784...72S}).
Meanwhile, the observed parameters in XTE\,J1751$-$305 can be only covered by the AIC channel for UCXBs in the mass-transfer rate versus the orbital period diagram (see Fig.\,9).

\section{BH-UCXBs}

\subsection{The observed BH-UCXB candidates}
It has been suggested that the mass-accretors in three UCXBs may be BHs, 
as follows:
(1) 47 Tuc X9 is a LMXB in the globular cluster 47 Tucanae. 
\citet{Miller-Jones2015MNRAS.453.3918M} first identified a radio counterpart to 47 Tuc X9,
suggesting that the ratio of radio/X-ray luminosity for 47 Tuc X9 is more consistent with BH-LMXBs 
and that the mass-donor might be a WD.
\citet{Bahramian2017MNRAS.467.2199B} recently reported the orbital period of 47 Tuc X9 as 
$28.18^{+0.02}_{-0.02}$\,min based on the Chandra data. They confirmed that 
this system is a BH-UCXB candidate in the Galaxy, suggesting that the mass-donor may be a CO WD.
In a further work, \citet{Tudor2018MNRAS.476.1889T} speculated that a BH+He star system could 
evolve into 47 Tuc X9.
(2) 
RZ 2109 is a globular cluster ultraluminous X-ray source in the elliptical galaxy NGC 4472.
It has been suggested that RZ 2109 is a BH-UCXB candidate with a WD donor
(see \citealt{Steele2014ApJ...785..147S,Dage2024MNRAS.529.1347D}).
(3)
M31 UCXB-1 is one of the brightest X-ray sources in the bulge of M31.
\cite{zhangJiachang2024MNRAS.530.2096Z} reported an orbital period of about 465\,s from this source
based on XMM-Newton and Chandra data.
\cite{MaQian-Qi2026} suggested that 
M31 UCXB-1 is a BH+WD system with the shortest orbital period among the known UCXBs.

\subsection{Formation channels}

The formation path to BH-UCXBs is still highly uncertain.
In globular clusters, dynamical interactions that pair BHs with companion stars are probably the main formation channel for BH-UCXBs (see, e.g., \citealt{Ivanova2005ApJ...621L.109I,XuanZeyuan2025ApJ...995...27X}). 
While in the Galactic field, similar to the formation of NS-UCXBs, there are also three classic formation channels, as follows: 
(1) The WD donor channel. 
According to detailed stellar evolution models with the convection- and rotation-boosted MB prescription,
\citet{YangXing-Peng2025ApJ...986..219Y} confirmed that 
isolated BH+MS binaries can evolve into BH-UCXBs with orbital periods of 7.7\,min as in M31 UCXB-1,
in which a low-mass WD transfers matter onto a BH.
Motivated by the fact that most massive stars are born in triple systems, 
\citet{XuanZeyuan2025ApJ...995...27X} argued that wide BH+WD systems 
can naturally form UCXBs via the eccentric Kozai-Lidov mechanism (see \citealt{Lidov1962P&SS....9..719L, Kozai1962AJ.....67..591K}).
(2) The He star donor channel.
\citet{QinKe2024ApJ...961..110Q} recently studied the formation and evolution of 
BH-UCXBs via the He star donor channel in a systematic way. 
Their simulations indicate that 
some newly formed BH+He star binaries after the CE stage could form UCXB-LISA sources 
with a maximum GW frequency $\sim5$\,mHz, which can be detected by LISA even in a distance of 100\,kpc. 
(3) The evolved MS donor channel. 
\citet{QinKe2023ApJ...944...83Q} suggested that
BH+MS binaries with initial orbital periods less than the bifurcation period could evolve into BH-UCXBs.
They predicted that LISA will detect a few BH-UCXB GW sources based on this channel.

For the formation of BH-LMXBs (the progenitors of BH-UCXBs), however, the progenitors of the BHs should be more massive. 
It is proven that the removal of the massive envelope of the BH progenitors via the CE evolution is very difficult in progenitor systems with low-mass ($1-2~M_\odot$) donor stars due to the large binding energy of the massive envelope (see, e.g., \citealt{pve97,kalo99,Podsiadlowski2003MNRAS.341..385P,ktls+16}). 
\cite{DengZhu-Ling2024ApJ...971...54D} 
argued that this difficulty can be circumvented if the BHs are formed through failed SNe, which allow BHs to be produced by stars with initial masses as low as $\sim17\,M_\odot$ (see \citealt{Horiuchi2014MNRAS.445L..99H}), thus avoiding the necessity of an overly massive progenitor and enabling the formation of BH‑LMXBs with compact.\footnote{The failed SN model 
is on the basis of the assumption that a BH is formed 
depending on the compactness of the stellar core at the moment of the collapse,  i.e.,
stellar cores with high compactness undergo failed SNe and form BHs, 
whereas stellar cores with low compactness produce NSs via SN explosions (see \citealt{Connor2011ApJ...730...70O}).}

As alternative channels, BH-LMXBs were proposed to evolve from BH binaries with intermediate-mass ($3-5~M_\odot$) donor stars by anomalous MB laws (see \citealt{Justham06}) or tidal torque between the system and a surrounding circumbinary disk (see \citealt{ChenWen-Cong2006MNRAS.373..305C}), but the simulated effective temperatures of donor stars are significantly higher than those of the observed values. If there exists a density spike of dark matter around the stellar-mass BH, \cite{QinKe2024ApJ...971...57Q} argued that the dynamical friction between dark matter and the mass-donors can drive the evolution of BH binaries into the observed BH-LMXBs, 
thus alleviating the effective temperature issue of BH-LMXBs with relatively 
short orbital periods.

Recently, \citet{ChenHai-Liang2023ApJ...951...91C} proposed that NS+WD binaries can evolve into BH-UCXBs
via the AIC of accreting NSs, provided the initial NS is both massive enough and able to accrete sufficient material, thus eventually collapsing into a BH.
For the signatures of the NS-AIC events,
it has been suggested that the magnetosphere of the NS  will be disrupted and
the surface of the NS would be hidden behind the event horizon of the BH during the collapse,
producing strong radio signatures (possibly observable as a nonrepeating fast radio burst; see \citealt{Falcke2014A&A...562A.137F}). 

In BH-UCXBs, the WD may enter the tidal radius of the BH and trigger a tidal disruption event (TDE) if the orbital shrinkage induced by the GW radiation is stronger than the orbital expansion caused by a mass-transfer process. \cite{YangXing-Peng2025ApJ...986..219Y} argued that a TDE may occur after 0.12\,Myr 
if M31 UCXB-1 is a BH-UCXB consisting of a $10~M_\odot$ BH and a $0.16~M_\odot$ WD.

\section{Compared with observations}

\subsection{The mass-transfer rate vs. the orbital period diagram}

\begin{figure}
\begin{center}
\epsfig{file=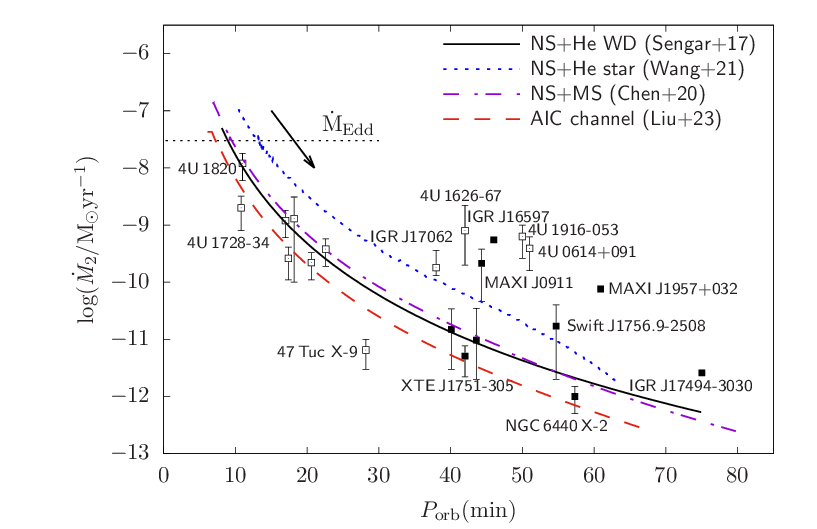,angle=0,width=11cm} 
\caption{Comparison between the theoretical evolutionary tracks from different channels 
and the observational data in the mass-transfer rate vs. the orbital period diagram. 
The black solid line shows the numerical relation fit based on the WD donor channel (see \citealt{Sengar2017MNRAS.470L...6S}),  the blue dotted line is based on the He star 
donor channel (see \citealt{Wang2021MNRAS.506.4654W}), the purple solid-dotted line is based on the 
evolved MS donor channel (see \citealt{ChenWen-Cong2020ApJ...900L...8C}) and the red dashed line is
based on the AIC channel (see \citealt{LiuD.2023MNRAS.521.6053L}).
The open squares and the solid squares denote persistent sources and transient sources in Table 1, respectively. Note that it is still controversial for the orbital periods of 4U 1728-34  (see \citealt{Heinke2013ApJ...768..184H}).
The dotted line is $\dot{\rm M}_{\rm Edd}$ for a NS accreting He-rich matter.}
\end{center}
\end{figure}

For different formation channels of UCXBs, the evolutionary tracks in the mass-transfer rate versus the orbital period diagram differ due to the distinct types or evolutionary states of their companions.
Thus, UCXBs could be identified by their locations in the mass-transfer rate versus 
the orbital period diagram (see Fig. 9).
As shown in Fig. 9, the majority of UCXBs can be well explained by the current existing models. 
Notably, some observed sources can only be accounted for by a specific formation channel (2-3 sources via the He star donor channel and 4-5 sources via the AIC channel), offering strong evidence for their progenitors. 
\citet{Heinke2013ApJ...768..184H} argued that the He star donor channel may 
explain the formation of the three wide-orbit persistent UCXBs with high mass-transfer rates.
However, it is still hard for any channels to form the three wide-orbit persistent sources. 
It is worth noting that recent studies on different MB models have offered new possibilities (e.g., \citealt{YangXing-Peng2025ApJ...995...99Y}). Meanwhile,   
the upper and lower error estimates for the three wide-orbit persistent sources are still not well determined (see \citealt{Heinke2013ApJ...768..184H}).

\subsection{The X-ray luminosity vs. the orbital period diagram}

\begin{figure}
\begin{center}
\epsfig{file=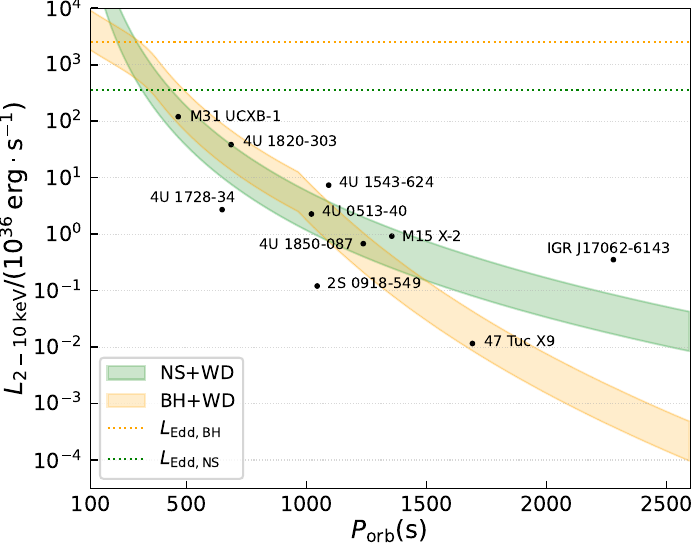,angle=0,width=11cm} 
\caption{Numerical results of the NS/BH+WD UCXB system in the $F_{\rm X}$[2-10\,keV] vs. $P_{\rm orb}$ diagram. In this figure, the observed persistent UCXBs (black solid dots) in Table 1 are presented  with the theoretical predictions for NS+WD (green shaded region) and BH+WD systems (orange shaded region). The theoretical results are obtained by assuming $M_{\rm NS}=1.4\,M_{\odot}$ and  $M_{\rm BH}=10\,M_{\odot}$. Source: from \citet{MaQian-Qi2026}. 
Compared with \citet{MaQian-Qi2026}, we added one source in this figure, i.e., 4U 1728-34.}
\end{center}
\end{figure}

Fig. 10 presents the comparison between the observed persistent UCXBs and the theoretical expectations in the X-ray luminosity  versus the orbital period diagram
 (see also \citealt{MaQian-Qi2026}).
The green and orange shaded regions denote the predicted relations for NS+WD and BH+WD binaries, computed assuming a NS mass of $1.4\,M_{\odot}$ and a BH mass of $10\,M_{\odot}$.
These theoretical bands are obtained from numerical models, in which the mass-transfer evolution of UCXBs is driven by the angular momentum losses due to the GW radiation
 (see, e.g., \citealt{Nelemans2010NewAR..54...87N, vanHaaften2012A&A...543A.121V}).
As GW emission becomes less efficient at wider orbit,
the predicted X-ray luminosity declines toward longer periods, which is
in agreement with the overall distribution of the persistent sources shown in this figure.

The numerical results indicate that, for $P_{\rm orb}\lesssim1000$\,s,
the predicted 2–10 keV luminosities of NS and BH accretors are nearly identical.
At longer periods ($P_{\rm orb}\gtrsim1000$\,s), however, the two evolutionary tracks separate.
As the accretion rate decreases with the increasing orbital period, NSs maintain an approximately constant radiative efficiency because of their solid surfaces, whereas BHs exhibit a decline in radiative efficiency at low accretion rates
 (see, e.g., \citealt{Narayan1995ApJ...452..710N, Watarai2001ApJ...549L..77W}).
This difference leads to a systematic divergence between the NS+WD and BH+WD predictions at longer orbital periods, enabling the two accretor types to be distinguished in this regime.
Among the systems displayed in Fig. 10,
only 47 Tuc X9 has been proposed as a BH-UCXB candidate
because of its high ratio of radio/X-ray luminosity. The location of 47 Tuc X9 in the luminosity versus the period diagram is consistent with the theoretical BH+WD relation, further supporting this interpretation.

\section{Impacts of UCXBs on some fields}

UCXBs are of great significance to modern astronomical research,
having important impacts on some astrophysical fields, for example, 
GW astronomy, multi-messenger astronomy,
binary evolution and the physics of NSs under extreme conditions, etc. 

\subsection{GW astronomy and multi-messenger astronomy}
One of the most prominent impacts of UCXBs lies in GW astronomy.
Current ground-based GW detectors operate primarily in the high-frequency band ($\gtrsim10$\,Hz),
making them sensitive to compact-object mergers such as double NSs and binary BHs  shortly before coalescence 
(see \citealt{Abbott2016PhRvL.116f1102A,AbbottPhysRevLett.119.161101}).
With orbital frequencies corresponding to GW frequencies of $\sim 10^{-4}-10^{-3}$\,Hz,
UCXBs emit nearly monochromatic and long-lived GW signals that fall squarely within the sensitivity band of planned space-based detectors such as LISA, TianQin and Taiji
(see, e.g., \citealt{Nelemans2010NewAR..54...87N,Tauris2018PhRvL.121m1105T,ChenWen-Cong2020ApJ...900L...8C,ChenHaiLiang2021MNRAS.503.3540C}).
\citet{YuShenghua2026MNRAS.545f2077Y} studied the post-contact evolution of Roche-lobe-filling NS+WD binaries 
and suggested that,
depending mainly on the WD mass and accretion efficiency, systems bifurcate into either stable UCXB-like mass-transfer or rapid orbital decay leading to merger.
They further demonstrated that these binaries produce diverse GW signals in the LISA band,
highlighting the role of the UCXB phase in producing resolvable LISA-band GW sources.
Previous studies indicate that the Milky Way may host on the order of several hundred UCXBs potentially detectable by future space-based GW missions 
(see, e.g., \citealt{ChenWen-Cong2020ApJ...900L...8C,Wang2021MNRAS.506.4654W,ChenMinghua2025ApJ...981..175C}).

Aside from their orbital GW emission in the mHz band,
UCXBs hosting rapidly rotating NSs may also produce high-frequency continuous GWs arising from stellar asymmetries or r-mode oscillations,
making them potential dual-band GW sources
(see, e.g., \citealt{Bildsten1998ApJ...501L..89B,ChenWen-Cong2021PhRvD.103j3004C}).
In such systems, the low-frequency orbital signal and the high-frequency 
spin-induced component could jointly constrain the properties of the NS, such as the mass, the radius, 
the moment of inertia and the ellipticity of the NS, etc. 
It has been proposed that 4U 1728-34 and 4U 1820-30 could be possible candidates 
for dual-band GW sources, being potentially detectable by aLIGO and LISA  simultaneously (see \citealt{Suvorov2021MNRAS.503.5495S}).

In addition to the GW emission, UCXBs are important multi-messenger systems.
Active mass-transfer powers persistent X-ray and often optical radiation 
from both the accretion flow and the irradiated donor star,
while systems that terminate mass-transfer can re-emerge as radio MSPs if the mass-accretors are NSs.
The combination of GW and electromagnetic observations provides complementary constraints on their orbital evolution and long-term accretion history.

\subsection{Binary evolution and NS physics}

Studies of UCXBs provide important constraints on key aspects of binary evolution, including the CE evolution, the angular-momentum loss (e.g., MB), and the  mass-transfer process, etc
(see, e.g., \citealt{Nelemans2010NewAR..54...87N, LinJie2018MNRAS.474.1922L, ChenHaiLiang2021MNRAS.503.3540C, Tauris2023pbse.book.....T, yang2024ApJ...974..298Y}).
They also offer valuable insights into NS physics, particularly the spin evolution and the maximum stable  masses for NSs
(see, e.g., \citealt{Tauris2012MNRAS.425.1601T,Kiziltan2013ApJ...778...66K,Alsing2018MNRAS.478.1377A}).

UCXBs are thought to originate from close binaries that have undergone one or two CE phases (see Figs 1, 3 and 7).
The CE phase is essential for shrinking the orbital separation from stellar scales to periods of order hours or less,
thereby enabling the subsequent evolution toward ultra-compact scale.
Thus, their orbital period distribution and donor masses provide valuable empirical constraints and tests for the CE theory.
In formation channels involving LMXBs, angular-momentum loss prior to the ultra-compact phase plays a critical role.
In particular, MB together with GW radiation governs the secular orbital shrinkage before and during RLOF.
\citet{Istrate2014A&A...571A..45I} showed that under conventional MB prescriptions,
it requires extreme fine-tuning of the initial donor mass and orbital period to form an UCXB from a detached MSP progenitor,
leading to a very narrow parameter space.
Subsequent studies demonstrated that the apparent fine-tuning issue 
is closely linked to incomplete understanding of MB laws
(see, e.g., \citealt{ChenHaiLiang2021MNRAS.503.3540C, DengZhuLing2021ApJ...909..174D}).
Alternative MB prescriptions or reduced efficiencies can significantly broaden the viable parameter space,
highlighting UCXBs as sensitive probes of angular-momentum loss mechanisms in low-mass binaries.

Most UCXBs host NSs with millisecond spin periods and very low-mass companions,
indicating long duration and stable mass-transfer
(see, e.g., \citealt{Alpar1982Natur.300..728A, Bhattacharya1991PhR...203....1B, ArmasPadilla2023AA...677A.186A}).
The long-term accretion phase naturally spins up the NS to millisecond periods and may substantially increase its mass,
establishing a direct evolutionary connection between UCXBs and radio MSPs.
The high incidence of accreting MSPs among UCXBs reinforces this link.
Because sustained accretion can push NSs toward their maximum mass,
UCXBs offer a potential avenue to constrain the NS equation of state
(see, e.g., \citealt{Kiziltan2013ApJ...778...66K,Alsing2018MNRAS.478.1377A}).
Moreover, the observed ratio between NS and BH accretors in ultra-compact systems 
provides important information on the high-mass end of the initial mass function of stars.

Observational evidence indicates
enrichment in different elements (e.g., He, C, O, Ne),
implying diversity in the chemical composition of the donor stars
(see \citealt{Juett2001ApJ...560L..59J, Koliopanos2021MNRAS.501..548K}).
The composition influences thermonuclear burst properties, thus offering an additional diagnostic of donor structure
(see \citealt{Cumming2001ApJ...557..958C}).
Explaining the observed chemical diversity requires donor stars that have reached different stages of nuclear burning and degrees of degeneracy, pointing to multiple formation pathways for UCXBs
(see, e.g., \citealt{Yungelson2002A&A...388..546Y, Nelemans2010NewAR..54...87N, Sengar2017MNRAS.470L...6S,ChenWen-Cong2020ApJ...900L...8C,Wang2021MNRAS.506.4654W}).

\section{Summary and perspective}

So far, about 50 UCXB candidates have been discovered, among which 
22 UCXBs have been confirmed with high confidence. 
It has been suggested  that most UCXBs may contain NS accretors,    
whereas three sources have been proposed to be BH-UCXB candidates 
though their nature is still undetermined. UCXBs play an important role in broad aspects of astrophysics.
They are expected to be promising sources of low-frequency GWs.
Their extremely short orbital periods place them in the mHz band,
making them potential targets for space-based GW detectors such as the LISA, TianQin and Taiji.

Currently, there are four evolutionary pathways that can lead to the formation of UCXBs, 
in which the WD donor channel is the main way for producing UCXBs. It remains uncertain for 
several key physical processes involved in the formation and evolution of UCXBs.
In particular, the efficiency of MB in low-mass donors is still not well constrained,
and different prescriptions for magnetic stellar winds can lead to significantly different orbital evolution and UCXB formation rates.
Another important uncertainty concerns the stability of mass-transfer in NS+WD binaries, depending sensitively on the structure of the mass-donor.
It also depends on the treatment of mass and angular momentum loss, 
including processes associated with the accretion disk such as disk winds and disk–orbit interactions.
These uncertainties in binary evolution modelling continue to limit robust predictions of the formation channels and population properties of UCXBs. 
It is worth noting that the detection of UCXBs through GWs will provide independent constraints on their orbital evolution and help to further test models of compact binary evolution.

Meanwhile, the chemical composition of UCXB donors remains a major unresolved issue 
and serves as a key evidence for distinguishing among different formation channels.
Distinct evolutionary pathways are expected to produce different surface element abundance
(see, e.g., \citealt{Yungelson2002A&A...388..546Y,Nelemans2010MNRAS.401.1347N}).
For example, the presence of $^{22}$Ne in the accretion disks of some UCXBs has been interpreted as possible evidence for the origin of the AIC channel.
Recently, \citet{Zhang2025ApJ...995L..64Z} reported the observations of the BW system PSR J2322–2650 using the James Webb Space Telescope (JWST).
This system has been suggested to originate from an UCXB with the He star donor channel
(see \citealt{GuoYun-Lang2022MNRAS.515.2725G}).
The observations reveal an extremely carbon-enriched atmosphere in the companion, with anomalously high C/O and C/N ratios.
These findings pose challenges to existing formation models, implying that the current formation channel may require additional physical processes to fully account for such abundances.
At present,  the 
observational constraints on the chemical composition of mass-donors in UCXBs remain very limited, but
will be greatly improved by current constraints by future high-sensitivity, multiwavelength spectroscopic observations with the ongoing and the next-generation large-aperture telescopes and advanced X-ray facilities, such as the JWST (see \citealt{JWST2006SSRv..123..485G}),
the Chinese Space Station Telescope (see \citealt{CSST2026SCPMA..6939501C}),  
the planned Extremely Large Telescopes (ELTs),
the Einstein Probe mission (see \citealt{YuanWeimin2025SCPMA..6839501Y}),
Athena (see \citealt{Nandra2013arXiv1306.2307N})
and Lynx (see \citealt{Gaskin2019JATIS...5b1001G}), etc.
Such efforts will enable systematic measurements of elemental abundances and help to understand the diverse formation pathways of UCXBs.

Additionally, the nature of the mass-accretors plays a crucial role in understanding the origin of UCXBs.
To date, nearly all confirmed UCXBs host NSs as the mass-accretors.
Most of them are accreting millisecond X-ray pulsars with spin periods in the millisecond range, indicating that they have undergone prolonged mass-transfer and have been efficiently spun up through mass-accretion.
Nevertheless,
systems containing more slowly rotating NSs also exist.
For example, 4U 1626-67 hosts a relatively slowly spinning NS ($7.68$\,s) 
and has been suggested to originate from the AIC channel. Alternatively,
the accretors in UCXBs can also be BHs.
Radio continuum observations provide a powerful tool for identifying BH accretors in UCXBs.
In particular, \citet{Miller-Jones2015MNRAS.453.3918M} proposed that 47 Tuc X9 is a strong BH-UCXB candidate based on its consistency with the radio/X-ray correlation for accreting BHs.
Its location in the orbital period–luminosity diagram  further supports this interpretation (see Fig. 10).
Future high-sensitivity radio facilities,
such as the Next Generation Very Large Array (ngVLA) and the Square Kilometre Array (SKA), will probe the radio properties of UCXBs to much deeper limits,
potentially revealing hidden BH-UCXBs and better constraining their formation and evolution.
In the future,
multi-wavelength observations of UCXBs are required, along with more samples of such sources and further theoretical simulations.

\begin{acknowledgements}
This study is supported by the National Natural Science Foundation of China (Nos 12225304, 12288102, 12273014, 12333008, 12422305 and 12403035), the CAS Project for Young Scientists in Basic Research (YSBR-148),
the National Key R\&D Program of China (No. 2021YFA1600404), the Yunnan Revitalization Talent Support Program (Young Talent Project and Yunling Scholar Project), the Yunnan Fundamental Research Project (Nos 202501AS070005 and 202605AS350010), and the International Centre of Supernovae (ICESUN), Yunnan Key Laboratory of Supernova Research (No. 202505AV340004).
\end{acknowledgements}

\bibliography{wang}
\bibliographystyle{raa}

\end{document}